\documentclass[a4paper,12pt]{article}
\usepackage{graphicx,epsfig,amsmath,amssymb}
\def\bma{\left( \begin{array} }
\def\ema{\end{array} \right)}
\newcommand{\vdel}{v_{\Delta}}
\newcommand{\delm}{\Delta M}

\begin{document}
\begin{flushright} KIAS-P12035  \end{flushright} 
\begin{center}
{\Large Same-Sign Tetra-Leptons from Type II Seesaw} 
\\ \vspace*{0.2in} 
{\large Eung Jin Chun and Pankaj Sharma} \\
\vspace*{0.2in}
{
Korea Institute for Advanced Study, Seoul 130-722, Korea
\rm }
\end{center}
\vspace*{0.6in}
\begin{center}
{\large\bf Abstract}
\end{center}

\noindent The type II seesaw mechanism introduces a hypercharged Higgs
triplet to explain the observed neutrino masses and mixing. Among
three triplet components, the doubly charged Higgs boson can be
the lightest and decay mainly to same-sign di-leptons.
Furthermore, the heavier singly charged or neutral Higgs boson
produces a doubly charged Higgs boson through its fast gauge
decay. This leads to a novel signature of same-sign tetra-leptons
resulting from a pair production of same-sign doubly charged Higgs
bosons caused by the nearly degenerate neutral scalar mixing which
can be sizable for an appropriate choice of the model parameters.
After studying production cross-sections for the same-sign tetra-lepton signal
in the parameter space of the mass splitting among triplet components and
the triplet vacuum expectation value,
we provide a LHC analysis of the same-sign tetra-lepton signal
for a benchmark point chosen to maximize the event number.

\vfill

\section{Introduction}

The origin of neutrino masses and mixing can be attributed to a
new ``Higgs triplet" which can couple to Higgs and lepton doublets
 and thus generate neutrino mass terms once it develops nontrivial
vacuum expectation value  (called the ``type II seesaw'' mechanism) \cite{magg}. One of the
distinctive features of the type II seesaw model is the presence
of a doubly charged Higgs boson which can be cleanly probed at
collider experiments \cite{gunion}--\cite{Muhlleitner:2003me}.
This allows an exciting possibility of discovering the neutrino
mass pattern at colliders by observing the lepton flavor structure
of the doubly charged Higgs boson decay to same-sign charged
lepton pairs \cite{Chun:2003ej}.  Because of such a nice property,
extensive studies at the LHC have been performed in the literature
\cite{Akeroyd:2005gt}--\cite{Akeroyd:2012nd}.

\medskip

In this paper, we investigate a novel signature of same-sign
tetra-leptons allowed in some parameter space of the type II
seesaw model that has not been studied so far. The mass splitting
among three Higgs triplet components (neutral, singly and doubly
charged) is controlled by a single coupling between the triplet
and the usual doublet Higgs bosons and its sign determines whether
the lightest component is a doubly charged or neutral one. When
the mass splitting is sizable and the doubly charged Higgs boson
is the lightest, the electroweak gauge interaction allows a fast
decay of the neutral or singly charged component of the Higgs
triplet into the lighter singly or doubly charged component.
Therefore, pair-produced Higgs triplet components can end up with
a pair of same-sign doubly charged Higgs bosons leading to
same-sign tetra-leptons if their leptonic Yukawa coupling is larger
than the ratio of the triplet and doublet Higgs
vacuum expectation values.

An essential feature of a  $l^\pm l^\pm l^\pm l^\pm$ signal
is that it occurs through the mixing between two nearly degenerate neutral scalars.
In the limit of vanishing triplet vacuum expectation value (VEV),
lepton number is conserved and thus there appear only $l^+ l^+ l^- l^-$ final states
coming from the production of $H^{++} H^{--}$ pairs  \cite{Akeroyd:2011zz,Akeroyd:2012nd}.
A non-vanishing triplet VEV arises due to the coupling between
the Higgs triplet and doublet which explicitly breaks lepton number in the Lagrangian.
This term  also induces  the
mass splitting between two real degrees of freedom in the neutral triplet scalar.
While this mass splitting becomes very small for a small triplet VEV,
the neutral triplet decay rate is also suppressed
by a small mass difference among the triplet components.
It turns out that there exists an optimal choice of the model parameters for which
the sizes of the neutral scalar mass splitting and decay rate become comparable and
the doubly charged Higgs boson decays mainly to same-sign di-lepton to
maximize the production cross-section of the $4l^\pm$ signal.

The same-sign tetra-lepton final state,
which is almost background free, provides an excellent new channel
to test the model and probe sizes of the Higgs triplet vacuum
expectation value and the mass splitting among the Higgs triplet
components at the LHC.\footnote{Let us note that
same-sign four leptons can also appear among others in the cascade decays of supersymmetric
particles if R-parity violating supersymmetry is assumed \cite{mukho11}.}

\medskip

In the next section, we will introduce the type II seesaw model
following the notation of Ref.~\cite{Chun:2003ej}. In Section III,
the branching ratios of the Higgs triplet components will be
studied in a parameter space of the mass splitting and the triplet
vacuum expectation value. Then, in Section IV analyzed are the
same-sign tetra-lepton signals at the LHC with two centre-of-mass energies
of $\sqrt{s}=8$ TeV (LHC8) and $\sqrt{s}=14$ TeV (LHC14) for the
doubly charged Higgs boson mass 400 GeV taking
a benchmark point in the two dimensional parameter space which maximizes
the the neutral boson mixing effect.  We conclude in Section V.

\section{The Type II Seesaw Model}

When the Higgs sector of the Standard Model is extended to have a
$Y=2$ complex $SU(2)_L$ scalar triplet $\Delta$ in addition to a
SM-Higgs doublet $\Phi$, the gauge-invariant Lagrangian is written
as
\begin{eqnarray}
\mathcal L=\nonumber\left(D_\mu\Phi\right)^\dagger
\left(D^\mu\Phi\right)  + \mbox{Tr}
\left(D_\mu\Delta\right)^\dagger\left(D^\mu\Delta\right) -\mathcal
L_Y - V(\Phi,\Delta)
\end{eqnarray}
where the leptonic part of the Lagrangian required to generate
neutrino masses is
\begin{equation} \label{leptonYuk}
\mathcal L_Y= f_{\alpha\beta}L_\alpha^T Ci\tau_2\Delta L_{\beta} +
\mbox{H.c.}
\end{equation}
and the scalar potential is
\begin{eqnarray}\label{Pot}
V(\Phi,\Delta)&=&\nonumber m^2\Phi^\dagger\Phi +
\lambda_1(\Phi^\dagger\Phi)^2+M^2\mbox{Tr}(\Delta^\dagger\Delta)\\\nonumber
&+&\lambda_2\left[\mbox{Tr}(\Delta^\dagger\Delta)\right]^2+\lambda_3\mbox{Det}(\Delta^\dagger\Delta)
+\lambda_4(\Phi^\dagger\Phi)\mbox{Tr}(\Delta^\dagger\Delta)\\
&+&\lambda_5(\Phi^\dagger\tau_i\Phi)\mbox{Tr}(\Delta^\dagger\tau_i\Delta)
+\left[\frac{1}{\sqrt{2}}\mu(\Phi^Ti\tau_2\Delta\Phi)+\mbox{H.c.}\right].
\end{eqnarray}
Here used is the $2\times 2$ matrix representation of $\Delta$:
\begin{equation}
\Delta=\bma{cc}
\Delta^+/\sqrt{2}  & \Delta^{++} \\
\Delta^0       & -\Delta^+/\sqrt{2} \ema .
\end{equation}
Upon the electroweak symmetry breaking with $\langle
\Phi^0\rangle=v_0/\sqrt{2}$, the $\mu$ term in Eq.~(\ref{Pot}) gives rise to the
vacuum expectation value of the triplet $\langle
\Delta^0\rangle=v_\Delta/\sqrt{2}$ where $\vdel\approx \mu
v_0^2/\sqrt{2}M^2$. We will assume $\mu$ is real positive without loss of generality.

After the electroweak symmetry breaking, there are seven physical
massive scalar eigenstates denoted by $H^{\pm,\pm}$, $H^\pm$,
$H^0$, $A^0$, $h^0$. Under the condition that $|\xi|\ll 1$ where
$\xi \equiv v_\Delta/v_0$, the first five states are mainly from
the triplet scalar and the last from the doublet scalar. For the
neutral pseudoscalar and charged scalar parts,
\begin{eqnarray}
 \phi^0_I = G^0 - 2 \xi A^0 \;, \qquad
 \phi^+ = G^+ + \sqrt{2} \xi H^+  \nonumber\\
 \Delta^0_I= A^0 + 2 \xi G^0 \;, \qquad
 \Delta^+= H^+ - \sqrt{2} \xi G^+
\end{eqnarray}
where $G^0$ and $G^+$ are the Goldstone modes, and
for the neutral scalar part,
\begin{eqnarray}
 \phi^0_R &=& h^0 - a \xi \, H^0 \,, \nonumber\\
 \Delta^0_R&=& H^0 + a \xi \, h^0
\end{eqnarray}
where $ a = 2 + 4 (4\lambda_1-\lambda_4-\lambda_5) M^2_{W}/
   g^2(M^2_{H^0}-M^2_{h^0}) $.
The masses of the Higgs bosons are
\begin{eqnarray} \label{massD}
 M^2_{H^{\pm\pm}} &=& M^2 + 2{\lambda_4 -\lambda_5 \over g^2 } M^2_{W}
 \nonumber\\
 M^2_{H^{\pm}} &=& M_{H^{\pm\pm}}^2 + 2{\lambda_5 \over g^2} M^2_{W}
 \nonumber\\
 M^2_{H^0, A^0} &=&  M^2_{H^\pm} +
    2{\lambda_5 \over g^2} M^2_{W} \,.
\end{eqnarray}
The mass of $h^0$ is
given by $m_{h^0}^2=4\lambda_1 v_\Phi^2$ as usual.

Eq.~(\ref{massD}) tells us that the mass splitting among triplet
scalars can be approximated as
\begin{equation}
 \delm \approx \frac{\lambda_5 M_W^2}{g^2 M} < M_W \,.
\end{equation}
Furthermore,  depending upon the sign of the coupling $\lambda_5$,
there are two mass hierarchies among the triplet components:
$M_{H^{\pm\pm}}>M_{H^\pm}>M_{H^0/A^0}$ for $\lambda_5<0$; or
$M_{H^{\pm\pm}}<M_{H^\pm}<M_{H^0/A^0}$ for $\lambda_5> 0$. In this
work, we focus on the latter scenario,  where the doubly charged
scalar $H^{\pm\pm}$ is the lightest so that it decays only to
$l_\alpha^\pm l_\beta^{\pm}$ or $W^\pm W^\pm$ whose coupling
constants are proportional to $f_{\alpha\beta}$ or $\xi$,
respectively. On the other hand, $H^0/A^0$  ($H^\pm$) decays
mainly to $H^\pm W^{\mp *}$ ($H^{\pm\pm} W^{\mp *}$) unless the
mass splitting $\Delta M$ is negligibly small. For more details,
see, e.g., Ref.~\cite{Chun:2003ej}.

An important quantity for a $4l^\pm$ signal is the
mass splitting $\delta M_{HA}$ between $H^0$ and $A^0$ which is much smaller than the mass difference
$\Delta M$ between different triplet components. The $\mu$ term in Eq.~(\ref{Pot}),
which is lepton number violating, generates not only the triplet VEV:
\begin{equation}
 v_{\Delta} = {\mu  v_0^2 \over \sqrt{2} M^2_{H^0}}\,,
\end{equation}
but also the mass splitting between the heavy neutral scalars,
$\delta M_{HA} \equiv M_{H^0} - M_{A^0}$:
\begin{equation} \label{MHA}
 \delta M_{HA} = 2 M_{H^0} {v_\Delta^2\over v_0^2}
 { M^2_{H^0} \over M^2_{H^0} - m^2_{h^0} } \,.
\end{equation}
As will be shown later, $\delta M_{HA}$ can be comparable to the total decay rate
of the neutral scalars, $\Gamma_{H^0/A^0}$, for a preferable choice of $v_\Delta$,
which enhances the same-sign tetra lepton signal.

Given the leptonic Yukawa term (\ref{leptonYuk}),
a non-zero triplet vacuum expectation value $\langle
\Delta^0\rangle$ gives rise to the neutrino mass matrix:
\begin{equation} \label{Mnu}
 M^\nu_{\alpha\beta} =  \sqrt{2} f_{\alpha\beta} v_\Delta
 \equiv U_{\alpha k} U_{\beta k} m_k e^{i \phi_k}
\end{equation}
where $U$ is the neutrino mixing matrix with Majorana phases $e^{i
\phi_k}$ factored out,  and $m_k$ are the neutrino mass
eigenvalues.  The mixing matrix $U$ is described by three angles
denoted by $\theta_{12}$, $\theta_{23}$, and $\theta_{13}$, and
one Dirac phase $\delta$.  The three angles and two mass-squared
differences $\Delta m^2_{21}$ and $\Delta m^2_{31}$ are fairly
well measured by various neutrino experiments \cite{forero12}. As
the sign of $\Delta m^2_{31}$ is not yet known, two neutrino mass
hierarchies are allowed. One is the normal hierarchy (NH) with
$m_3>m_2>m_1$ and the other is the inverted hierarchy (IH) with
$m_2>m_1>m_3$. Thus, the neutrino mass matrix (\ref{Mnu}) can be
fully reconstructed after some assumptions on values of the CP
phases, $\delta$ and $\phi_k$, and one neutrino mass, say $m_1$
($m_3$) for NH (IH) which can be as small as zero.

For the collider analysis, we will take benchmark points for
each neutrino mass hierarchy assuming $m_1 (m_3)=0$ for NH (IH)
and vanishing CP phases. Taking the best fit values
\cite{forero12} given by $\Delta m^2_{21}=7.62\times10^{-5}
\,\mbox{eV}^2$, $\Delta m^2_{31}=2.53\times10^{-3} (-2.40)
\,\mbox{eV}^2$, $\sin^2\theta_{12} = 0.320$,
$\sin^2\theta_{23}=0.49 (0.53)$ and $\sin^2\theta_{13} = 0.026
(0.027)$ for NH (IH), we get the following neutrino mass matrix in
the eV unit:
\begin{equation} \label{MnuNH}
 M^\nu = \left(\begin{array}{ccc} 0.00403 & 0.00816 &
0.00259 \cr 0.00816 & 0.0264 & 0.0215 \cr 0.00259 & 0.0215 &
0.0286 \cr\end{array}\right) \; \mbox{for NH}
\end{equation}
and
\begin{equation} \label{MnuIH}
M^\nu = \left(\begin{array}{ccc} 0.0479 & -0.00557 & -0.00573\cr
-0.00557 &  0.0239 & -0.0240 \cr -0.00573 & -0.0240 & 0.02693 \cr
\end{array}\right) \; \mbox{for IH}.
\end{equation}
Thus, the leptonic Yukawa coupling $f$ can be obtained from
Eq.~(\ref{Mnu}) given the triplet vacuum expectation value
$v_\Delta$. To get an estimate of the overall size of $f$ it is useful to remember
$f v_\Delta \sim 10^{-2}$ eV derived from a rough relation  of $f v_\Delta \sim m_k$.
Recall that the neutrino mass matrices (\ref{MnuNH},\ref{MnuIH}) tell about the
flavor-dependent branching fractions of $H^{\pm\pm}$ and thus observation of
$ee/e\mu/\mu\mu$ final states at the LHC will be able to determine which
neutrino mass pattern is right \cite{Chun:2003ej}.

\section{Phase diagrams for branching fractions of Triplet decay}

Apart from the parameters set by the neutrino data in the coupling
$f$, there are two more free parameters in the type II seesaw
model: the triplet mass splitting $\Delta M$ and the triplet vacuum
expectation value $v_\Delta$. Depending on these, the triplet
components have different decay properties.
In Table.~\ref{decay}, we show the possible decay channels for the
triplet scalars for the case where $H^{++}$ is the lightest.  The
decays to the lepton and  quark/di-boson final states are due to
the Yukawa coupling $f$ and a mixture of the doublet and triplet
Higgs controlled by $\xi=v_\Delta/v_0$, respectively. On the other
hand, the final states with off-shell $W$ is due to the usual
$SU(2)$ gauge interaction which dominates if allowed
kinematically.

\begin{table}[h]
\begin{center}
\begin{tabular}{|l|l|l|l|}
\hline $~~~~~~H^0$ &   $~~~~~~A^0$ & $~~~~~~H^+$ & $~~~~~H^{++}$
\\\hline
$\to t\bar t$       &   $\to t\bar t$           &   $\to t\bar b$       &   $\to \ell^+ \ell^+ $\\
$\to b\bar b$       &   $\to b\bar b$           &   $\to \ell^+\nu$     &   $\to W^{+} W^{+} $\\
$\to \nu\bar{\nu}$  &   $\to \nu\bar{\nu}$      &   $\to W^+Z$      &   \\
$\to ZZ $       &   $\to Zh^0$          &   $\to W^+h^0$        &   \\
$\to h^0h^0$   &   $\to H^{\pm}W^{\mp^*}$      &   $\to H^{++}W^{-^*}$ &   \\
$\to  H^{\pm}W^{\mp^*}$     &                   & &   \\\hline
\end{tabular}\caption{\label{decay}Possible decay channels for the triplet Higgs bosons for $\lambda_5>0$.}
\end{center}
\end{table}

\begin{figure}[h]
\begin{center}
\includegraphics[width=55mm]{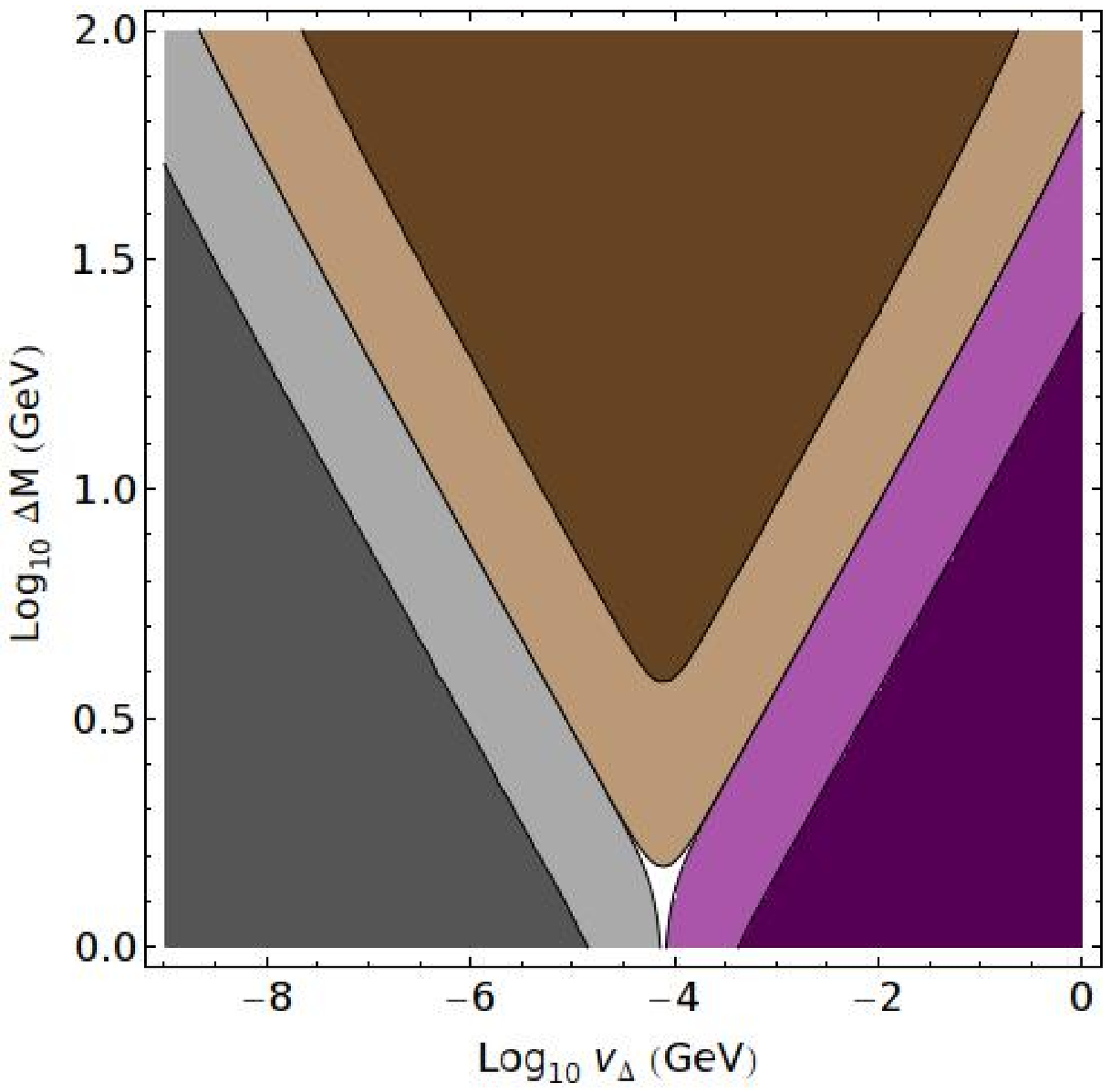}
\includegraphics[width=55mm]{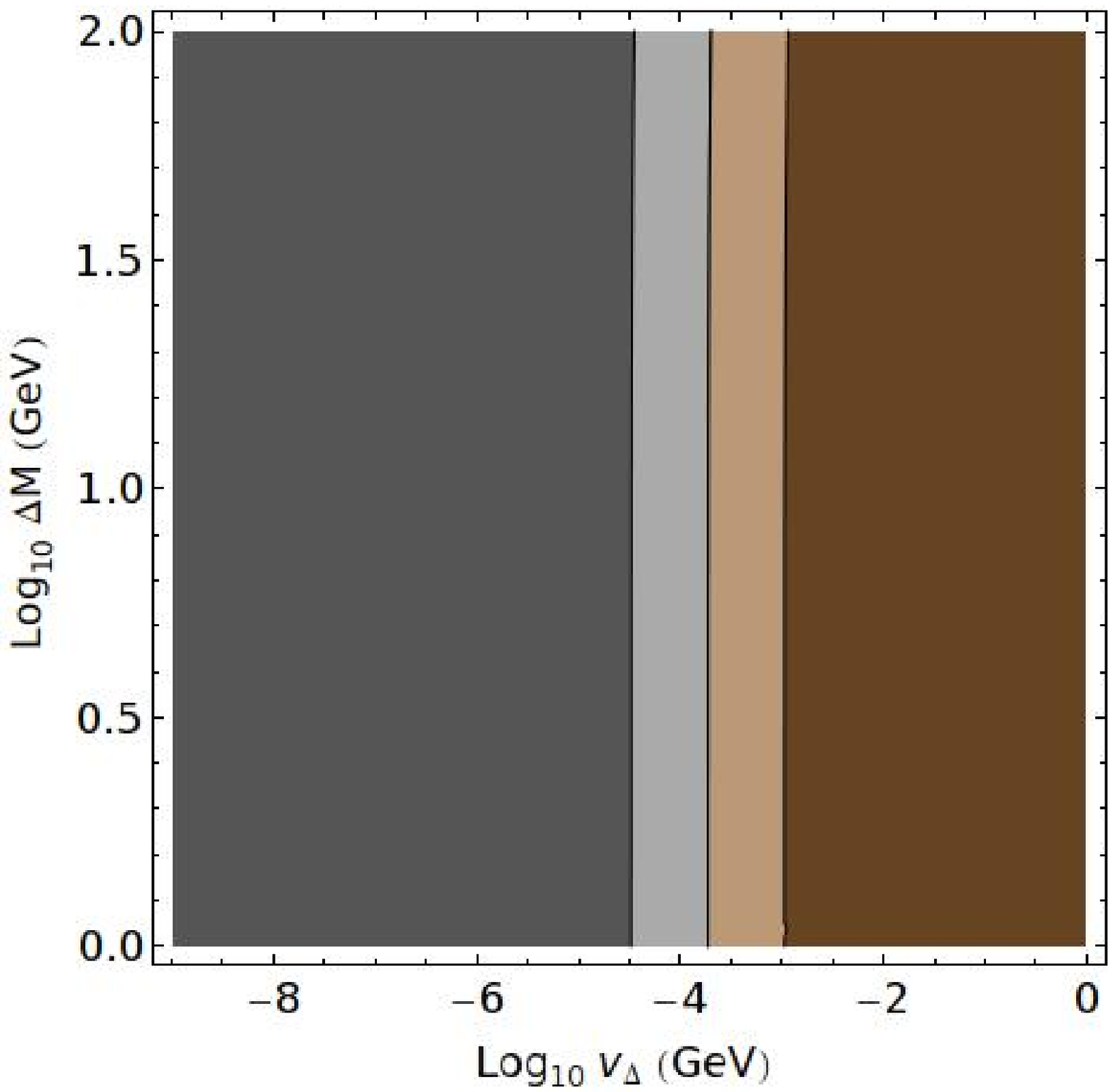}
\end{center}
\caption{ Phase diagrams for $H^+$(left) and $H^{++}$(right)
decays in the type II seesaw model for $\lambda_5>0$ with
$M_H^{++}=300$ GeV. The dark-colored regions denote the branching
fraction larger than 99\%.}\label{decay:Hp}
\end{figure}

Summing over all the lepton final states in the triplet decays,
the flavor structure of neutrino mass matrix does not matter as
the total leptonic decay rate is proportional to $\sum_k m_k^2$.
Keeping this parameter to be (0.1 eV)$^2$, we calculate the
branching fractions (BFs) for each scalar depending on two
parameters: $\Delta M$ and $v_\Delta$.  In Fig.~\ref{decay:Hp}, we
show phase diagrams for $H^+$ and $H^{++}$ decays in the plane of
$\Delta M$ and $\vdel$. In the left panel, the brown, the gray
and the purple regions show the branching fractions for the decays
$H^+\to H^{++}W^{-^*}$, $H^+\to\ell^+\nu$ and $H^+\to \{t\bar
b,W^+Z,W^+h\},$ respectively. In the right panel, the brown and
the gray regions show the branching fractions for the decays
$H^{++}\to W^+W^+$ and $H^{++}\to \ell^+\ell^+$ respectively. In
both panels, the dark-colored regions denote the parameter space
where the branching fraction is greater than 99\% and the
light-colored regions denote the parameter space where the
branching fraction is between 50\%-99\%.

As it can be seen from Fig.~\ref{decay:Hp}, the leptonic decay of
$H^+$ ($H^+\to\ell^+\nu$) is dominant for small values of
$v_\Delta<0.1$ MeV (corresponding to $|f| > |\xi|$).  For low
values of $v_\Delta$ and $\Delta M$, this BF is always greater
than 0.99. However, for moderate mass splitting $\delm>5$ GeV, the
decay $H^+\to H^{++}W^{-^*}$ start becoming dominant at
$v_\Delta=0.1$ MeV. For slightly larger value of $\Delta M$ at
around 10-20 GeV, there is much larger parameter space opening up
for this decay. Hence, we can see that for moderate $v_\Delta$
(around 1 keV-10 MeV), large mass splitting is allowed for the
branching fraction larger than 99\%. On the other hand, for large
$v_\Delta$ and low mass splitting, rest of the decays viz.,
$(H^+\to t\bar b, W^+Z, W^+ h)$ have appreciable contributions and
thus BF($H^+\to H^{++}W^{-^*})$ goes down.

For the $H^{\pm\pm}$ decays, we see that for $v_\Delta<0.1$ MeV,
it is completely dominated by leptonic decay i.e., $H^{\pm\pm}\to
\ell^\pm\ell^\pm$ while for $v_\Delta>1$ MeV, it is dominated by
the decay to two $W^\pm$s. These BFs are completely independent of
the mass splitting $\Delta M$ as is obvious from that the
$H^{\pm\pm}$ is the lightest.

\begin{figure}[h]
\begin{center}
\includegraphics[width=55mm]{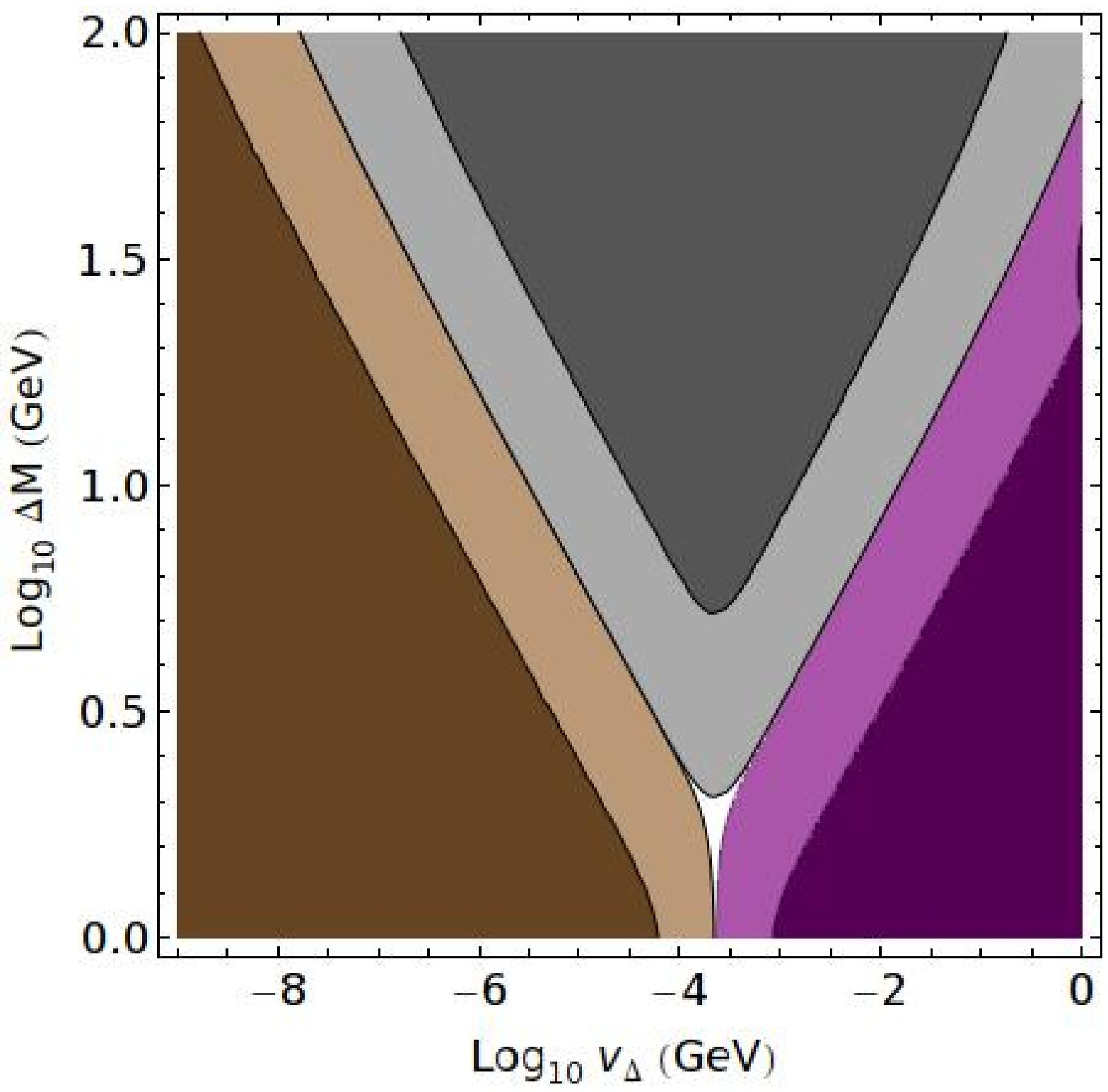}
\includegraphics[width=55mm]{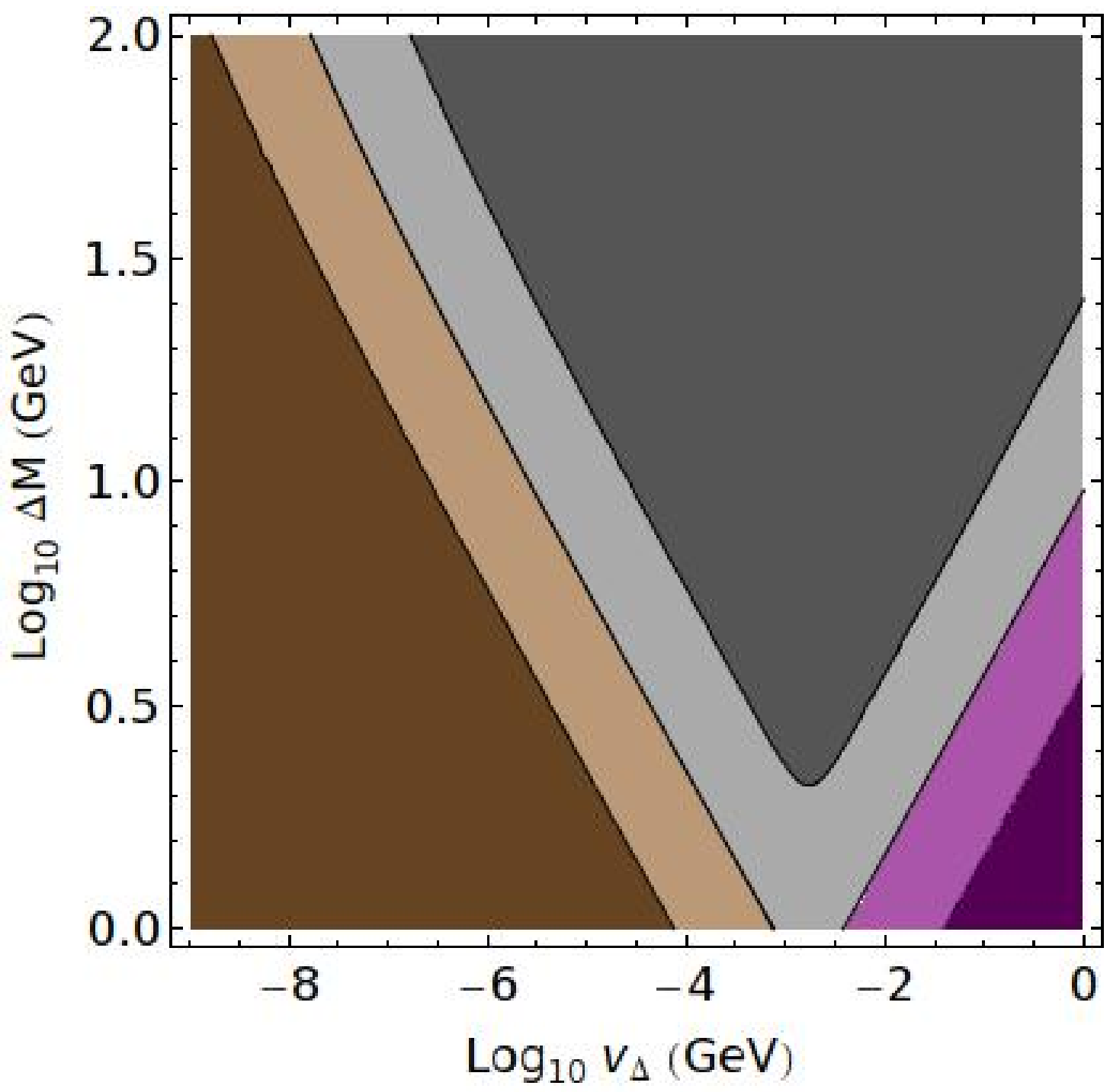}
\end{center}
\caption{ Phase diagrams for $H^0$(left) and $A^0$(right)  decays
in the type II seesaw model for $\lambda_5>0$ with $M_H^{++}=300$
GeV. The dark-colored regions denote the branching fraction in the
range of 49\%-50\%.}\label{decay:H0}
\end{figure}

Similarly, in Fig.~\ref{decay:H0}, we show phase diagrams  for
$H^0$ and $A^0$ decays in the plane of $\delm$ and $\vdel$. In the
left panel, the brown, the gray and the purple regions show the
BFs for the decays $H^0\to\nu\bar{\nu}$, $H^0\to H^{+}W^{-^*}$,
and $H^0\to \{t\bar t, b\bar b, ZZ,h^0h^0\}$ respectively. In the
right panel, the brown, the gray and the purple regions show the
BFs for the decays $A^0\to\nu\bar{\nu}$, $A^0\to H^{+}W^{-^*}$,
and $A^0\to \{t\bar t, b\bar b, Zh^0\}$ respectively. In both
panels, the dark-colored regions denote the parameter space where
BF is between 49\%-50\% and the light-colored regions denote the
parameter space where the BF is between 20\%-49\%.

\section{Same-Sign Tetra-Leptons at the LHC}

A remarkable feature of the Type II seesaw model would be the
observation of doubly charged Higgs bosons $H^{\pm\pm}$ at the TeV
scale. Production of such scalars has been extensively studied in
processes $q\bar q\to Z^*/\gamma^*\to H^{++}H^{--}$
\cite{Muhlleitner:2003me,Han:2007bk,delAguila:2008cj,Akeroyd:2011zz}
and $q^\prime\bar q\to W^*\to H^{\pm\pm}H^{\mp}$
\cite{Akeroyd:2005gt}. The pair production mechanism leads to four
lepton signals including a pair of same-sign di-leptons with
opposite charges, or tri-lepton signals which are relatively clean and
thus studied by many authors.
In this section, we study a new possibility of
probing the type II seesaw model at the LHC through a distinctive
signal of four same-sign leptons which are either positively or
negatively charged.  The processes which contribute to such a
signal are as follows:
\begin{enumerate}
\item\label{pro:1} $q^\prime \bar q\to W^*\to H^{\pm}H^0/A^0$\\
proceeded by $H^\pm\to H^{\pm\pm}W^{\mp^*}$  and $H^0/A^0\to
H^{\pm}W^{\mp^*}\to H^{\pm\pm}W^{\mp^*}W^{\mp^*}$;
\item\label{pro:3} $q \bar q\to Z^*\to H^0A^0$\\
proceeded by  $H^0/A^0\to H^{\pm}W^{\mp^*}\to
H^{\pm\pm}W^{\mp^*}W^{\mp^*}$.
\end{enumerate}
The above mentioned production cross-sections of triplet scalars
only depend on the scalar masses because the interactions are due
to the triplet gauge couplings. The cross-section for process
\ref{pro:3} is the largest among the three. Out of all the triplet
pairs produced in the above three processes, only some fraction of them
eventually give same-sign $H^{\pm\pm}$ pairs whose production is controlled
by the neutral scalar mixing parameter as will be discussed below.

We further assume leptonic decays of $H^{\pm\pm}$ to be dominant,
which is allowed for a large part of parameter space (with $|f|> |\xi|$)
 as can be
seen from Fig.~\ref{decay:Hp}. The processes \ref{pro:1}
and \ref{pro:3} give a signal which contains $4\ell^\pm + X$ where $X$ contains
jets and leptons coming from off-shell $W$s which are soft and thus hard to be detected
due to a small mass gap between the triplet components.

As noted in Refs.~\cite{Akeroyd:2011zz,Akeroyd:2012nd},
the $4l^\pm$ final state cannot occur in the limit of lepton number conservation, that is,
$\mu, v_\Delta \to 0$, due to the cancelling interference between $H^0$ and $A^0$.
However, the above processes 1 and 2 are allowed when there is a finite mass difference
(\ref{MHA}) violating lepton number.
In the limit of $M_{H^0/A^0} \gg \delta M_{HA}, \Gamma_{H^0/A^0}$
as well as $\Gamma_{H^0} \simeq \Gamma_{A^0}$, one can find a simple expression for the $4l^\pm$ production rate following a proper treatment of
the interference effect in the narrow width approximation.
Note that this phenomenon occurs as $H^0$ and $A^0$,
sharing the same final states, can mix together like in the $B^0$-$\bar{B}^0$ system.
In other words, $\Delta^0$ can oscillate to $\Delta^{0\dagger}$ and vice versa to produce
wrong sign leptons in our system.
If they undergo sufficient oscillation before they decay, i.e., $\delta M_{HA} \gtrsim \Gamma_{H^0/A^0}$,
the lepton number violating production of same-sign tetra-leptons becomes sizable.  That is,
this effect is controlled by the usual oscillation parameter $x_{HA}$:
\begin{equation}
 x_{HA} \equiv { \delta\! M_{HA} \over \Gamma_{H^0/A^0} } \,.
\end{equation}
From the calculation of the $H^0$-$A^0$ interference term in the narrow width approximation, we obtain the cross-sections for the processes \ref{pro:1} and \ref{pro:3} as follows:
\begin{eqnarray}
\sigma\left(4\ell^\pm + 3W^{\mp^*}\right)&=& \sigma\left(pp\to
H^\pm H^0 + H^\pm A^0\right)  \left[{ x_{HA}^2\over 1+ x_{HA}^2}\right]
 \mbox{BF}(H^0/A^0\to H^\pm W^{\mp^*}) \nonumber\\
&\times& \left[\mbox{BF}(H^\pm\to H^{\pm\pm} W^{\mp^*})\right]^2
\left[\mbox{BF}(H^{\pm\pm}\to \ell^\pm\ell^\pm)\right]^2;
\label{4l3W}\\
\sigma\left(4\ell^\pm + 4W^{\mp^*}\right)&=& \sigma\left(pp\to
H^0A^0\right) \left[{2+x^2_{HA} \over 1+x^2_{HA}}
{ x_{HA}^2\over 1+ x_{HA}^2} \right]
 \mbox{BF}(H^0\to H^\pm W^{\mp^*})\nonumber\\
 &\times&\mbox{BF}(A^0\to H^\pm W^{\mp^*})
 \left[ \mbox{BF}(H^\pm\to H^{\pm\pm}W^{\mp^*})\right]^2
\left[\mbox{BF}(H^{\pm\pm}\to \ell^\pm\ell^\pm)\right]^2.
 \label{4l4W}
\end{eqnarray}
As expected, the cross-sections vanish for $x_{HA} \to 0$
 recovering the lepton number conserving limit. In the limit of $x_{HA}\gg1$ (the maximal
 lepton number violation), the
 mixing factors become one and the $4l^\pm$ signal numbers are controlled only by
 the branching fractions of $H^0$ and $A^0$.
 For a rough estimation of
$x_{HA}$, let us compare
 $\delta M_{HA}$ in Eq.~(\ref{MHA}) with the gauge decay rate
of $H^0/A^0$ given by
\begin{equation}
 \Gamma_{H^0/A^0} \sim {G_F^2 \Delta M^5\over \pi^3}.
\end{equation}
From this one finds that $x_{HA}\sim 1$ can be obtained with, e.g.,
$v_\Delta \sim 10^{-4}$ GeV and $\Delta M \sim 2$ GeV.

\begin{figure}[t]
 \begin{center}
 \includegraphics[width=55mm,angle=-90]{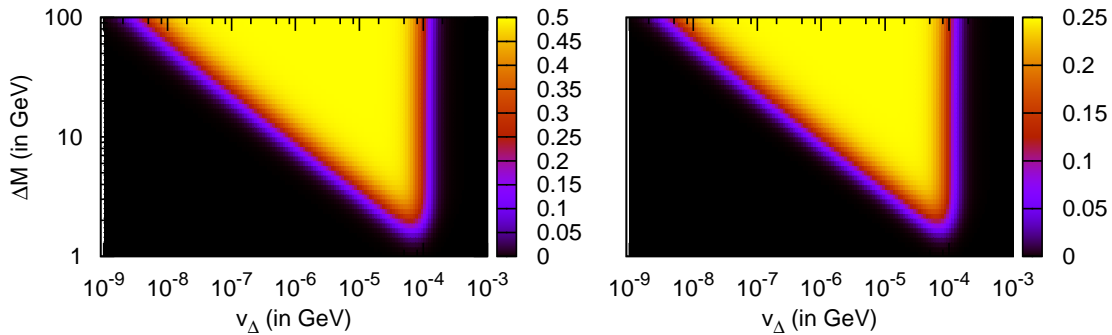}
 \end{center}
 \caption{Product of branching fractions for processes \ref{pro:1}
  (left), and \ref{pro:3} (right).
 \label{BFprod}}
 \end{figure}

\begin{figure}[t]
\begin{center}
 \includegraphics[width=90mm,angle=-90]{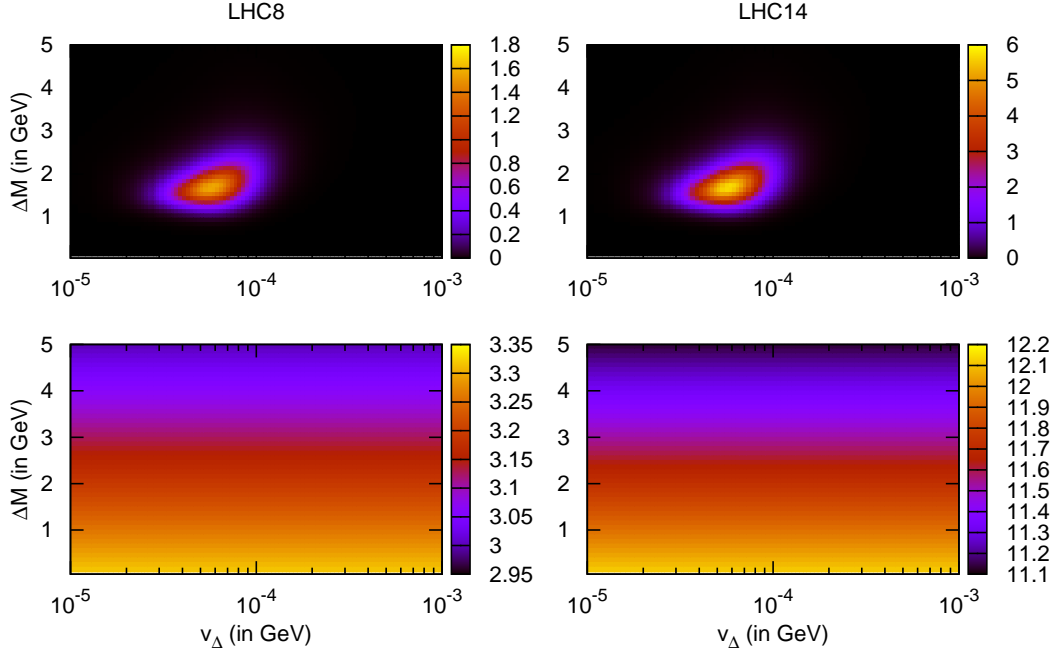}
\end{center}
\caption{Bottom panels show  cross-sections (in fb) for $H^\pm
H^0/H^\pm A^0/H^0A^0$ production, and top panels for same-sign
tetra-lepton events. Left (right) panels are for LHC8 (LHC14).
 \label{comb-cross}}
\end{figure}

Since we are interested in four same-sign leptons in the final
state, the cross-section for such a signal also depend on $\vdel$
through the decay branching fractions. These signals depend on
[BF]$^5$ and [BF]$^6$, so we should look at those regions of
parameter space where these BFs are maximum. To explore those
regions in the $\Delta M- v_\Delta$ plane, we plot the products of
BFs which occur in the evaluation of cross-sections of those
signals. In Fig.~\ref{BFprod}, we show product of BFs for
processes \ref{pro:1}  (left figure) and for process \ref{pro:3}
(right figure) in $\Delta M-v_\Delta$ plane. One can see from
these figures that there are large parameter space where these
products are maximum i.e., $0.5$. In Fig.~\ref{comb-cross}
(bottom), we show sum of cross-sections for processes \ref{pro:1}
and \ref{pro:3} which can finally give same-sign tetra-leptons.
The cross-sections are independent of $v_\Delta$ and steadily
decreases with the rise of $\Delta M$. In Fig.~\ref{comb-cross}
(top), we show cross-sections for
$\ell^\pm\ell^\pm\ell^\pm\ell^\pm$ signal at LHC8 and LHC14 in the
$\Delta M-v_\Delta$ plane. The Fig.~\ref{comb-cross} (top) is
obtained by superposing the Figs.~\ref{comb-cross} (bottom) and
\ref{BFprod} and multiplying with the oscillation factor as in
Eqs.~(\ref{4l3W},\ref{4l4W}).   The doubly charged Higgs mass is
taken to be 400 GeV. Note that the CMS experiment puts a lower
bound on the doubly charged Higgs boson which decays only to
charged leptons: $M_{H^{\pm\pm}}> (330-360)$ GeV considering the
pair production only \cite{cms}.  One can see from
Fig.~\ref{comb-cross} that the same-sign tetra-lepton
cross-section is maximized for $v_\Delta = (10^{-4} - 10^{-5})$
GeV and $\Delta M=(1-2)$ GeV in accordance with the rough
estimate. To recapitulate the condition for large number of same-sign tetra-lepton 
events, one needs $x_{HA} \gtrsim 1$ enhancing the $H^0$-$A^0$ mixing 
and appropriate $\Delta M$ small enough to increase the production of $H^0/A^0 H^\pm$, {\it etc},
but not too small to suppress the decays $H^0/A^0 \to H^\pm W^*$, {\it etc}, assuming
of course the dominance of the leptonic decays channels of $H^{\pm\pm}$.
As far as the latter condition is realized, the type II seesaw can be tested by observing
the usual signals of a pair of same-sign di-leptons from $H^{++} H^{--} \to l^+ l^+ l^- l^-$.
If the parameter region sits near the limited bright region in Fig.~\ref{comb-cross},
one can look in addition for four same-sign leptons allowing 
to get information about $v_\Delta$ and $\Delta M$. Furthermore,
in the limit of $x_{HA} \gg 1$, the transition probability of $\Delta^0$ to $\Delta^{0\dagger}$
is maximized to be one-half rendering the number of same-sign tetra-leptons comparable to that 
of a pair of same-sign di-leptons, and thus it will be more efficient to look for the same-sign tetra-lepton signals.

\medskip

 We remark that lepton flavor violating processes like $\mu \to e e\bar{e}$ are highly 
suppressed in the parameter region of our interest. For $v_\Delta \sim (10^{-4}-10^{-5})$ GeV, the 
neutrino mass relation (\ref{Mnu}) requires $f_{ij} \sim (10^{-6}-10^{-7})$, whereas one needs, for instance,
$f_{11}f_{12} > 10^{-8}$ to observe the signal of $\mu \to e e\bar{e}$ in the future experiments
\cite{Chun:2003ej}. 

\medskip

Now let us take a benchmark point with $\vdel=7\times10^{-5}$ GeV,
$\Delta M=1.5$ GeV and $M_{H^\pm\pm}=400$ GeV which  gives $\delta
M_{HA} =3.68 \times 10^{-11}$ GeV, $\Gamma_{H^0/A^0} =3.73\times
10^{-11}$ GeV, and thus $x^2_{HA}/(1+x_{HA}^2) =0.79$.
 In Table.~\ref{crossHH}, we show the
values of pair production cross-sections relevant for our
calculation at LHC8 and LHC14.


So far, we have not distinguished among flavors of charged leptons
i.e.,  $e, \mu, \tau$ in our analysis. However, we know that at
the LHC, $\tau$ leptons are more difficult to identify and have
large backgrounds. $\tau$ leptons can decay leptonically $\tau\to
e\nu_e\bar \nu_\tau$ and $\tau\to \mu\nu_\mu\bar \nu_\tau$ with
branching fractions of 17\% each. These $e$'s and $\mu$'s are less
energetic than their parent $\tau$'s. On the other hand, decays
$H^{\pm,\pm}\to e^\pm e^\pm/\mu^\pm\mu^\pm/e^\pm\mu^\pm$ are much
cleaner and produce the two energetic $e$ and $\mu$ closer to
invariant mass $M_{H^{\pm\pm}}$. For the collider analysis
including lepton flavor dependence, we consider the full neutrino
mass matrices calculated in Eqs.~(\ref{MnuNH}, \ref{MnuIH}) for
the NH and IH, respectively. In the NH case, the BF of the
$H^{\pm\pm}$ decay to the $e$ and $\mu$ final states is 32\%, and
thus $H^{\pm\pm}$  decay mainly to same-sign $\tau$ pairs. Of
course, the leptonic decays of these $\tau$'s to $e/\mu$ are
included in our analysis. On the other hand, for the IH case, the $H^{\pm\pm}$
decay BF to the $e$ and $\mu$ final states is 60\% and thus we
expect to have more same-sign tetra-lepton signal events compared
to the NH case.

\begin{table}
\begin{center}
\begin{tabular}{|c|c|c|}\hline
Final State      &    $\sigma/$fb (8 TeV)     &$\sigma/$fb (14
TeV)
\\\hline $H^+H^0$    &    0.761                & 2.931
\\\hline $H^+A^0$    &    0.761                & 2.931
\\\hline $H^-H^0$    &    0.275                & 1.209
\\\hline $H^-A^0$    &    0.275                & 1.209
\\\hline $H^0A^0$    &    1.014                & 4.322
\\\hline
\end{tabular}\caption{\label{crossHH}The values of cross-section
for different sub-processes contributing to same-sign tetra-lepton
signals for the LHC8 and LHC14. We use $\Delta M=1.5$ GeV and
$M_{H^{\pm\pm}}=400$ GeV for LHC8 and LHC14.}
\end{center}
\end{table}

In Table \ref{events}, we show number of events before selection
cuts and after selection cuts for both NH and IH at two center of
mass energies 8 TeV and 14 TeV LHC. We assume 15 fb$^{-1}$ and 100
fb$^{-1}$ of integrated luminosities for LHC8 and LHC14
respectively.


We expect almost no background to our same-sign tetra-lepton
signal. The only potential background can come from multi $W$
production with at least four $W^\pm$ and extra $W^\mp$ demanding
four $W^\pm$ decaying leptonically and the rest $W^\mp$
hardronically. In the lowest order, there is a diagram for   $4
W^\pm +2 W^\mp$ production whose  cross-section is proportional to
$\alpha_{EW}^7$. Other background can come from $gg\to 4W^++4W^-$
which is a loop process with top- and bottom-quarks inside the
loop. The cross-section for this process would be suppressed by
$\alpha_S^2\alpha_{EW}^8$ times loop-suppression factor. Thus, the
background for resulting number of same-sign tetra-lepton final
states is practically zero at the LHC.

Since there is negligible background, the selection  criteria for
the leptons are very trivial. We just need to have those cuts
which are essential for detector acceptance regions. So, the basic
cuts like $p_T>20$ GeV and $|\eta|<$2.5 for all leptons would be
sufficient to detect our signal.
We use \texttt{CTEQ6L} \cite{Pumplin:2005rh} parton  distribution
function (PDF) and the renormalization/factorization scale is set
at $2M_{H^+}$. We use \texttt{CALCHEP} \cite{Pukhov:2004ca} to
generate the parton level events for the relevant processes. Then,
using \texttt{LHEF} \cite{Alwall:2006yp} interface, we pass these
parton level events to \texttt{PYTHIA} \cite{Sjostrand:2006za} for
fragmentation and initial/final state radiations. We use
\texttt{PYCELL}, a toy calorimeter in \texttt{PYTHIA}, for
hadronic level simulation for finding jets using a cone algorithm.
For realistic simulation, we use the following criteria for selection of events:
\begin{itemize}
\item There should be exactly 4 isolated leptons ($e,\mu$) of same
sign in the events, \item $p_T^\ell>20$ GeV, $|\eta|<2.5$ and
$\Delta R(\ell_i,\ell_j)>0.2$. \item leptons are arranged in
decreasing order in $p_T$ and are labelled as $\ell_i$,
($i=1,\ldots,4$) in that order. \item no jet should overlap with
any lepton.
\end{itemize}

In Table \ref{events} we show the total number of four same-sign
lepton events before and after applying selection cuts. We list
these numbers for both NH and IH scenarios for both LHC8 and LHC14
with 15 fb$^{-1}$ and 100 fb$^{-1}$ of integrated luminosities
respectively. We find that after applying the section cuts, the
total number of events are reduced by 25\%. The number of events
in the IH scenario is twice the number of events in NH scenario.
This is due to the fact that the branching ratio
$\mbox{BF}(H^{\pm\pm}\to ee/e\mu/\mu\mu)$ is almost twice in IH
scenario relative to NH as noted before. One can see that the
signal event numbers are large enough to test the type II seesaw
mechanism through same-sign tetra-lepton final states.

The same-sign tetra-lepton signal for $M_{H^{++}}=400$ GeV might
be barely observable at LHC8 for the IH case, but the event number
is too small to reconstruct its mass.   On the other hand, LHC14
with 100 fb$^{-1}$ of integrated luminosity would have large
number of events to look for the doubly charged Higgs mass of
$M_{H^{++}}=400$ GeV.
 Assuming that 10 signal events
would be sufficient for the claim of discovery, we also find that
$H^{\pm\pm}$ with mass $M_{H^{\pm\pm}}$ as large as  600 GeV and
700 GeV can be probed for NH and IH scenario respectively at the
LHC14 with 100 fb$^{-1}$ of integrated luminosity.

\begin{table}[h]
\begin{center}
\begin{tabular}{l|c|c}
\hline
                                &   Pre-selection   &   Selection\\
\hline
$\ell^\pm\ell^\pm\ell^\pm\ell^\pm ~~ \mbox{(LHC8-NH)}$  &   4      &   3  \\
$\ell^\pm\ell^\pm\ell^\pm\ell^\pm ~~ \mbox{(LHC8-IH)}$  &  9 & 8 \\
\hline
$\ell^\pm\ell^\pm\ell^\pm\ell^\pm ~ \mbox{(LHC14-NH)}$   &   110     &   94  \\
$\ell^\pm\ell^\pm\ell^\pm\ell^\pm ~ \mbox{(LHC14-IH)}$   &   240     &   210 \\
\hline
\end{tabular}\caption{\label{events}
Number  of events for same-sign tetra-lepton signals before and after selection cuts for both NH and IH
scenarios at LHC8 and LHC14 with 15 fb$^{-1}$ and 100 fb$^{-1}$ of integrated luminosities respectively.}
\end{center}
\end{table}

\begin{figure}[h]
\begin{center}
\includegraphics[scale=0.65,angle=-90]{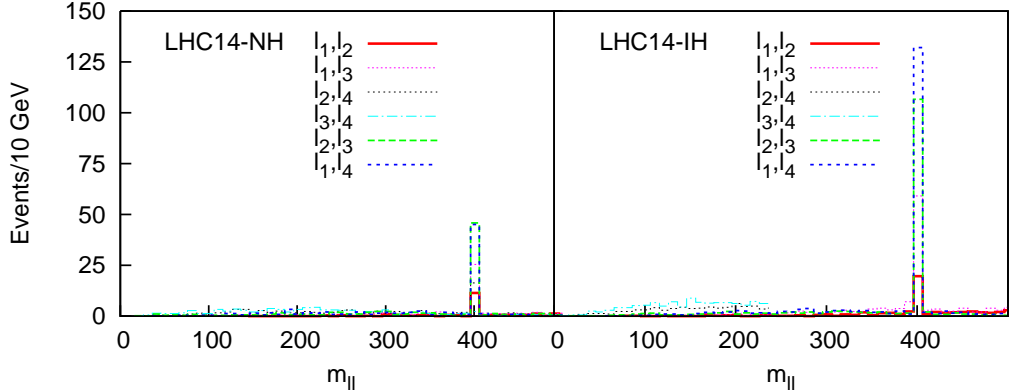}
\end{center}
\caption{Invariant mass from 4$\ell^\pm$ final states.\label{InvMass}}
\end{figure}

In Fig.~\ref{InvMass}, we plot the reconstructed  $H^{\pm\pm}$
mass from the sample of selected same-sign tetra-lepton events for
both NH and IH neutrino mass scenarios at LHC14 with 100 fb$^{-1}$
integrated luminosity. The peaks in all plots correspond to
$H^{\pm\pm}\to ee/e\mu/\mu\mu$ decays while the broad part
(off-peak) of distribution correspond to $\tau$ decays. For IH,
the number of events at the peak is about 2.5 times larger than NH
and the peaks are more pronounced.  One can clearly see that the
doubly charged Higgs boson mass can be readily found also from the
same-sign tetra-lepton signal for our benchmark point.

\section{Conclusion}

It is pointed out that a remarkable signal of  same-sign tetra-lepton,
 $l^\pm l^\pm l^\pm l^\pm$, is allowed in the type II seesaw mechanism
 introducing a Higgs triplet
as the origin of neutrino masses and mixing. Observability of such a signal at the LHC
strongly depends on the mass splitting $\Delta M$ among the triplet components and
the triplet vacuum expectation value $v_{\Delta}$ (or the  overall size of
the leptonic Yukawa coupling given by $f \sim 0.01 \mbox{eV}/ v_\Delta$).
When the doubly charged component $H^{\pm\pm}$ is the lightest,
larger $\Delta M$ allows more efficient gauge decay of the neutral component
to the singly charged one and then to the doubly charged  one. Thus, a pair production
of the triplet components at colliders can end up with producing $H^{\pm\pm}H^{\pm\pm}$
whose branching fraction to same-sign tetra-leptons becomes larger for smaller $v_\Delta$.
Another crucial ingredient for increasing the $4l^\pm$ signal number is the $H^0$-$A^0$ mixing
parameter $x_{HA}$ which becomes smaller for smaller $v_\Delta$ and larger $\Delta M$.
Therefore, there appear optimal values of the model parameters which maximize the same-sign
tetra-lepton signal.

After studying such a behavior in the $\Delta M-v_\Delta$ plane, we identified a benchmark point
with $\Delta M=1.5$ GeV and $v_\Delta = 7\times10^{-5}$ GeV for
maximized the signal numbers at the LHC.
The collider analysis of a same-sign tetra-lepton signal is trivial as it is completely background
free. After making the typical selection cuts to identify four same-sign leptons, we have shown that
one can obtain sizable event numbers for $M_{H^{\pm\pm}} =400$  GeV with integrated luminosity
of 100 fb$^{-1}$ at the LHC14 to reconstruct the doubly charged Higgs mass
in both cases of the normal and inverted neutrino mass hierarchy.
We also found that the doubly charged Higgs boson mass
up to  600 GeV and 700 GeV can be probed at LHC14 for NH and IH scenarios for our benchmark point,
assuming 10 signal events for a discovery of the type II seesaw mechanism.

With accumulating data at the LHC, it is worthwhile to make a discovery or exclusion study
of the type II seesaw mechanism in the full parameter space of
$M_{H^{\pm\pm}}$, $\Delta M$ and $v_{\Delta}$ through the $l^\pm l^\pm l^\pm l^\pm$ signal
 as well as
the conventional signal of $l^+ l^+ l^- l^-$
followed by the production of $H^{++} H^{--}$ which has been studied extensively in the literature.
We leave this as future work.

\medskip
{\bf Acknowledgments:}
We are grateful to Andrew Akeroyd and Hiroaki Sugiyama for their comments on
the interference effect in the narrow width approximation.
EJC was supported by the National Research Foundation of Korea (NRF) grant funded by the Korea government (MEST) (No.~20120001177).


\begin{thebibliography}{50}

\def\plb#1#2#3{Phys.\ Lett.\       {\bf B#1}  (#2) #3}
\def\npb#1#2#3{Nucl.\ Phys.\       {\bf B#1}  (#2) #3}
\def\prd#1#2#3{Phys.\ Rev.\        {\bf D#1}  (#2) #3}
\def\prl#1#2#3{Phys.\ Rev.\ Lett.\ {\bf #1}   (#2) #3}
\def\mpl#1#2#3{Mod.\ Phys.\ Lett.\ {\bf A#1}  (#2) #3}
\def\rep#1#2#3{Phys.\ Rep.\        {\bf #1}   (#2) #3}
\def\sci#1#2#3{Science             {\bf #1}   (#2) #3}
\def\astro#1#2#3{Astrophys.\ J.\   {\bf #1}   (#2) #3}
\def\epj#1#2#3{Eur.\ Phys.\ J.\   {\bf C#1}   (#2) #3}
\def\jhep#1#2#3{JHEP              {\bf #1}   (#2) #3}
\def\ptp#1#2#3{Prog.\ Theor.\ Phys.\ {\bf #1}  (#2) #3}

\bibitem{magg}
  M.~Magg and C.~Wetterich,
  Phys.\ Lett.\ B {\bf 94} (1980) 61;
%
  T.~P.~Cheng and L.~-F.~Li,
  Phys.\ Rev.\ D {\bf 22} (1980) 2860;
  R.~N.~Mohapatra and G.~Senjanovic,
  Phys.\ Rev.\ D {\bf 23} (1981) 165.


\bibitem{gunion}
 J.F. Gunion, J. Grifols, A. Mendez, B. Kayser and F. Olness,
 \prd{40}{1989}{1546};
 R. Vega and D. Dicus, \npb{329}{1990}{533};
 J.F. Gunion, R. Vega and J. Wudka, \prd{42}{1990}{1673};
\bibitem{godbole}
 R. Godbole, B. Mukhopadhyaya and M. Nowakowski, \plb{352}{1995}{388};
 K. Cheung, R. Phillips and A. Pilaftsis, \prd{51}{1995}{4731};
 K. Huitu, J. Maalampi, A. Pietila and M. Raidal, \npb{487}{1997}{27}.
\bibitem{rizzo}
 T.G. Rizzo, \prd{45}{1992}{42};
 N. Lepore, B. Thorndyke, H. Nadeau and D. London, \prd{50}{1994}{2031}.
\bibitem{dion}
 B. Dion {\it et. al.}, \prd{59}{1999}{075006};
 A. Datta and A. Raychaudhuri, \prd{62}{2000}{055002}.
\bibitem{ma}
 E. Ma, M. Raidal and U. Sarkar, \prl{85}{2000}{3769}; \npb{615}{2001}{313}.


\bibitem{Muhlleitner:2003me}
  M.~Muhlleitner and M.~Spira,
  Phys.\ Rev.\ D {\bf 68}, 117701 (2003)
  [hep-ph/0305288].


\bibitem{Chun:2003ej}
  E.~J.~Chun, K.~Y.~Lee and S.~C.~Park,
  Phys.\ Lett.\ B {\bf 566} (2003) 142
  [hep-ph/0304069].


\bibitem{Akeroyd:2005gt}
  A.~G.~Akeroyd and M.~Aoki,
  Phys.\ Rev.\ D {\bf 72} (2005) 035011
  [hep-ph/0506176].

\bibitem{Han:2007bk}
  T.~Han, B.~Mukhopadhyaya, Z.~Si and K.~Wang,
  Phys.\ Rev.\ D {\bf 76} (2007) 075013
  [arXiv:0706.0441 [hep-ph]].

\bibitem{Garayoa:2007fw}
  J.~Garayoa and T.~Schwetz,
  JHEP {\bf 0803} (2008) 009
  [arXiv:0712.1453 [hep-ph]].


\bibitem{Kadastik:2007yd}
  M.~Kadastik, M.~Raidal and L.~Rebane,
  Phys.\ Rev.\ D {\bf 77} (2008) 115023
  [arXiv:0712.3912 [hep-ph]].

\bibitem{Akeroyd:2007zv}
  A.~G.~Akeroyd, M.~Aoki and H.~Sugiyama,
  Phys.\ Rev.\ D {\bf 77} (2008) 075010
  [arXiv:0712.4019 [hep-ph]].


\bibitem{Perez:2008ha}
  P.~Fileviez Perez, T.~Han, G.~-y.~Huang, T.~Li and K.~Wang,
  Phys.\ Rev.\ D {\bf 78} (2008) 015018
  [arXiv:0805.3536 [hep-ph]].

\bibitem{delAguila:2008cj}
  F.~del Aguila and J.~A.~Aguilar-Saavedra,
  Nucl.\ Phys.\ B {\bf 813}, 22 (2009)
  [arXiv:0808.2468 [hep-ph]].

\bibitem{Akeroyd:2010ip}
  A.~G.~Akeroyd, C.~-W.~Chiang and N.~Gaur,
  JHEP {\bf 1011} (2010) 005
  [arXiv:1009.2780 [hep-ph]].

\bibitem{Akeroyd:2011zz}
  A.~G.~Akeroyd and H.~Sugiyama,
  Phys.\ Rev.\ D {\bf 84} (2011) 035010
  [arXiv:1105.2209 [hep-ph]].

\bibitem{Melfo:2011nx}
  A.~Melfo, M.~Nemevsek, F.~Nesti, G.~Senjanovic and Y.~Zhang,
  Phys.\ Rev.\ D {\bf 85}             (2012) 055018
  [arXiv:1108.4416 [hep-ph]].

\bibitem{Aoki:2011pz}
  M.~Aoki, S.~Kanemura and K.~Yagyu,
  Phys.\ Rev.\ D {\bf 85} (2012) 055007
  [arXiv:1110.4625 [hep-ph]].

\bibitem{Akeroyd:2012nd}
  A.~G.~Akeroyd, S.~Moretti and H.~Sugiyama,
  Phys.\ Rev.\ D {\bf 85} (2012) 055026
  [arXiv:1201.5047 [hep-ph]].

\bibitem{mukho11}
  B.~Mukhopadhyaya and S.~Mukhopadhyay,
  Phys.\ Rev.\ D {\bf 82}             (2010) 031501
  [arXiv:1005.3051 [hep-ph]];
  Phys.\ Rev.\ D {\bf 84}             (2011) 095001
  [arXiv:1108.4921 [hep-ph]].

\bibitem{forero12}
For a recent update, see,
  D.~V.~Forero, M.~Tortola and J.~W.~F.~Valle,
  arXiv:1205.4018 [hep-ph].

\bibitem{cms}
CMS Collaboration, CMS-PAS-HIG-12-005

\bibitem{Pumplin:2005rh}
  J.~Pumplin, A.~Belyaev, J.~Huston, D.~Stump and W.~K.~Tung,
  JHEP {\bf 0602}, 032 (2006)
  [hep-ph/0512167].
\bibitem{Pukhov:2004ca}
  A.~Pukhov,
  hep-ph/0412191.

\bibitem{Alwall:2006yp}
  J.~Alwall, A.~Ballestrero, P.~Bartalini, S.~Belov, E.~Boos, A.~Buckley, J.~M.~Butterworth and L.~Dudko {\it et al.},
  Comput.\ Phys.\ Commun.\  {\bf 176}, 300 (2007)
  [hep-ph/0609017].

\bibitem{Sjostrand:2006za}
  T.~Sjostrand, S.~Mrenna and P.~Z.~Skands,
  JHEP {\bf 0605}, 026 (2006)
  [hep-ph/0603175].



\end{thebibliography}
\end{document}